\documentclass[aps,twocolumn,prb,superscriptaddress]{revtex4}
\usepackage[pdftex]{color,graphicx}
\usepackage{dcolumn}
\usepackage{amsmath}
\usepackage{natbib}

\begin{document}

\title{Lattice and spin excitations in multiferroic h-YbMnO$_3$}

\author{J. Liu}
\author{C. Toulouse}
\affiliation{Laboratoire Mat\'eriaux et Ph\'enom\`enes Quantiques UMR 7162 CNRS, Universit\'e Paris Diderot-Paris 7, 75205 Paris cedex 13, France}
\author{P. Rovillain}
\affiliation{Laboratoire Mat\'eriaux et Ph\'enom\`enes Quantiques UMR 7162 CNRS, Universit\'e Paris Diderot-Paris 7, 75205 Paris cedex 13, France}
\affiliation{School of Physics, University of New South Wales, Sydney, New South Wales 2052, Australia}
\affiliation{The Bragg Institute, ANSTO, Kirrawee DC NSW 2234, Australia}
\author{M. Cazayous}
\author{Y. Gallais}
\author{M-A. Measson}
\affiliation{Laboratoire Mat\'eriaux et Ph\'enom\`enes Quantiques UMR 7162 CNRS, Universit\'e Paris Diderot-Paris 7, 75205 Paris cedex 13, France}
\author{N. Lee}
\author{S. W. Cheong}
\affiliation{Rutgers Center for Emergent Materials and Department of Physics and Astronomy, Rutgers University, 136 Frelinghuysen Road, Piscataway, NJ 08854, USA}
\author{A. Sacuto}
\affiliation{Laboratoire Mat\'eriaux et Ph\'enom\`enes Quantiques UMR 7162 CNRS, Universit\'e Paris Diderot-Paris 7, 75205 Paris cedex 13, France}

\date{\today}     

\begin{abstract}
Lattice and spin excitations have been studied by Raman scattering in hexagonal YbMnO$_3$ single crystals. 
The temperature dependences of the phonon modes show that the E$_2$ mode at 256 cm$^{-1}$ related to the displacement of Mn and O ions in a-b plane 
is coupled to the spin order. The $A_1$ phonon mode at 678 cm$^{-1}$ presents a soft mode behavior at the N\'eel temperature. Connected to the motion of the apical oxygen ions along the c direction,
this mode controls directly the Mn-Mn interactions between adjacent Mn planes and the superexchange path. 
Crystal field and magnon mode excitations have been identified. The temperature investigation of the spin excitations shows that the spin structure is strongly influence by the Yb-Mn interaction. 
Under a magnetic field along the c axis, we have investigated the magnetic reordering and its impact on the spin excitations.  
\end{abstract}


\maketitle

\section{Introduction}

Multiferroics have aroused great attention for the past years following the discovery of compounds that display giant cross-coupling effects between magnetic and ferroelectric order parameters. These materials are of great technological significance. The multifunctional applications of these compounds include magnetoelectric memory storage, electric-field control of magnetic sensors, and ferroelectrics based field-effect transistors.\cite{Yang2007, Fiebig2002, Moussa1996, Tokura2006} Furthermore, due to the associated rich magnetic-electric-elastic phase diagram, the interactions between the two orders  are also under intense investigations.\cite{Eerenstein2006, Chu2008, Baek2010}

In magnetic multiferroic materials, the ferroelectric order is induced by particular magnetic structures. The RMnO$_3$ rare-earth manganites are one of the most investigated family of this type of multiferroics. These compounds crystallize in orthorhombic structures for larger ionic radius (R= La, Ce, Pr, Nd, Sm, Eu, Gd, Tb, Dy) or  in hexagonal structure (space group : P6$_3$cm ) for R with smaller ionic radius (R= Ho, Er, Tm, Yb, Lu, Y).\cite{Fiebig2005, Cheong2007, Park2003} The hexagonal manganites exhibit ferroelectric order at high temperatures and antiferromagnetic order at low temperatures.\cite{Lee2008, Pimenov2006}
 In the orthorhombic case, the magnetic frustrations lead to spin-lattice coupling which is induced by the inverse Dzyaloshinski-Moriya interaction. Whereas in the hexagonal case, as YbMnO$_3$ studied in this work, the tilting of MnO$_5$ bipyramids and buckling of rare-earth layers are responsible for multiferroicity.\cite{Aken2004}
However, the nature of the interactions which control the magnetic structure in h-RMnO$_3$ is still unclear. For example, the Mn moments are perpendicular to the a and b axes, and their arrangement in the upper plane is either antiparallel or parallel in h-HoMnO$_3$ and h-YbMnO$_3$, respectively. The stability of a magnetic configuration rather than an other and the nature of the spin waves have to be understood.
The magnetic structure of polycrystalline h-YbMnO$_3$ samples have been investigated by infrared (IR)\cite{Divis2008}, inelastic neutron scattering\cite{Fabreges2008}, M$\ddot{o}$ssbauer spectroscopy\cite{Fabreges2009} and second harmonic generation (SHG)\cite{Fiebig2003}, and the lattice vibrations have been studied in h-YbMnO$_3$ epitaxial film and polycrystal samples.\cite{Fukumura2007, Fukumura2009} However, measurements on YbMnO$_3$ single crystals are scarce.

In this article, we present polarized Raman spectra of h-YbMnO$_3$ and the A$_1$, $E_1$ and $E_2$ phonon modes. The temperature dependence of the A$_1$ modes along the $c$ direction shows a renormalization of the 678 cm$^{-1}$ phonon mode at the magnetic transition. This mode related to displacement of the apical oxygen ions modulates the Mn-Mn interaction between the adjacent Mn planes. 
The magnetic excitations and the magnetoelectric phase diagram have also been investigated. We have identified the spin excitations and we show that the Mn spin orientation in the (a,b) plane is strongly influenced by the Yb-O-Mn interaction. The measurements under a magnetic field indicates that the interplane interaction control the magnetic transition. 

\par

\section{Experimental Details}

\par

YbMnO$_3$ single crystals are grown using the high-temperature flux growth technique.\cite{Kim2000} The samples are millimeter size platelets with the hexagonal c-axis perpendicular to the surface and a thickness of about 0.1 mm.  The crystals have been polished to obtain high surface quality for optical measurements.    
Raman spectra were recorded in a backscattering geometry with a triple spectrometer Jobin Yvon T64000 coupled to a liquid-nitrogen-cooled CCD detector using the 514 and 647 nm excitation line from a Ar$^+$-Kr$^+$ mixed gas laser. The high rejection rate of the spectrometer allows us to detect the magnons at frequencies lower than 50 cm$^{-1}$. Our resolution of the excitation frequencies is about 0.75~cm$^{-1}$. Measurements between 7 and 300 K have been performed using an ARS closed-cycle He cryostat. The measurements under a magnetic field up to 10 T have been obtained using an Oxford Spectromag split-coil magnet. 

\section{Results and discussion}

\subsection{Lattice excitations}

\begin{figure}
 \includegraphics*[width=8cm]{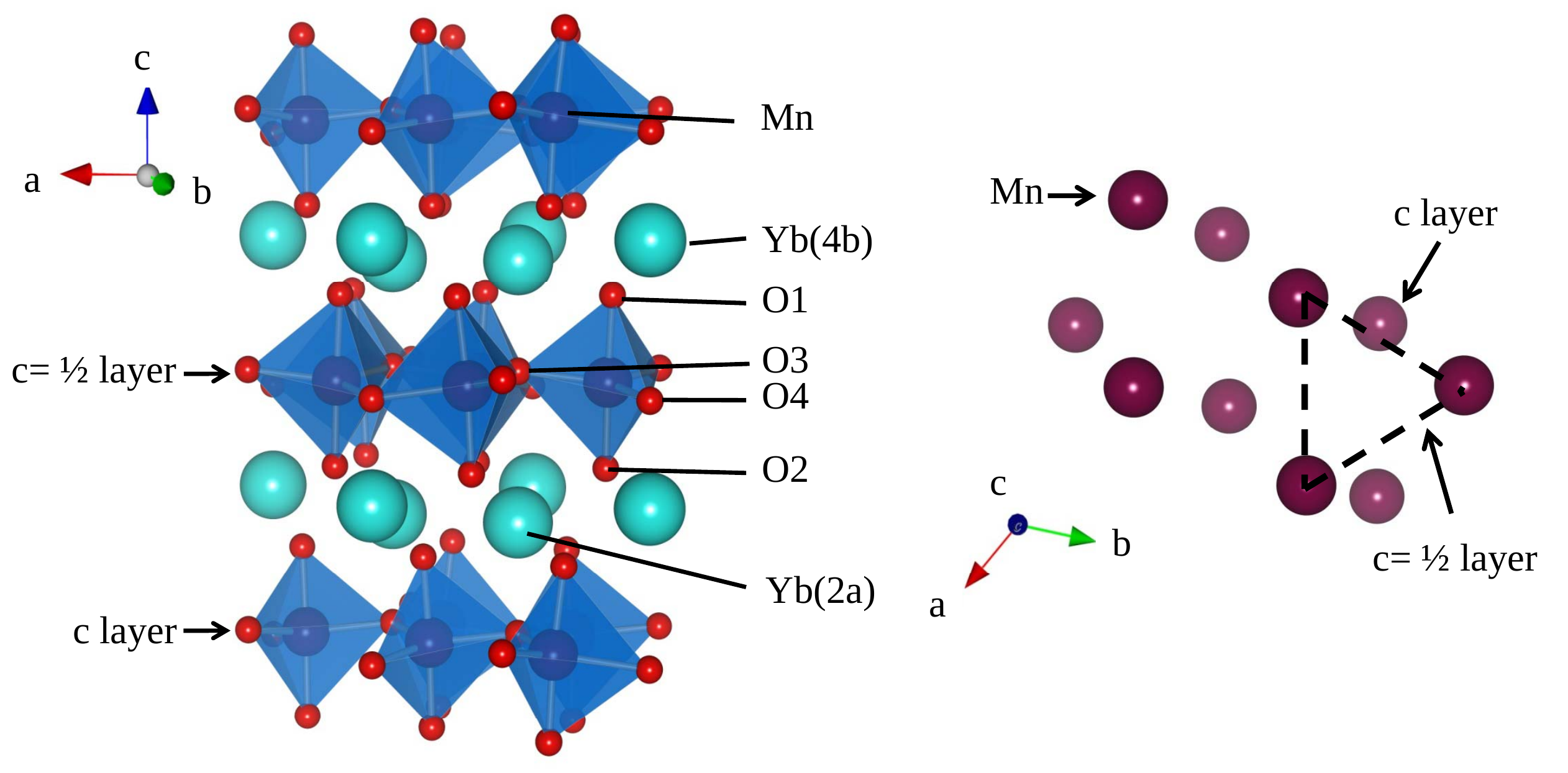} 
 \caption{\label{Fig1} 	 
 Structure of hexagonal YbMnO$_3$ in its ferroelectric P6$_3$cm phase.}
\end{figure}

YbMnO$_3$ crystallizes in the hexagonal space group P6$_3$cm. The unit cell of the ferroelectric structure is shown in Fig. \ref{Fig1}.
The lattice constants at 300 K are a = 6.0629(1) $\dot{A}$ and c = 11.3529(1) $\dot{A}$. 
YbMnO$_3$ is formed of tilted MnO$_5$ bipyramids arranged in a layered type structure in the a-b plane with apical (O$_1$, O$_2$) and in-plane (O$_3$, O$_4$) oxygen ions. Between the bipyramid layers, the rare-earth ions layers are stacked along c axis.\cite{Munoz2000} These MnO$_5$ bipyramids are two-dimensionally connected with each other at their corners and formed the triangular lattice of the Mn$^{3+}$ ions. The displacement of ions along the c axis induced a ferroelectric transition around 900 K.   
Mn$^{3+}$ spins order antiferromagnetically in the ab plane below the N\'eel temperature T$_N$=80~K. In the RMnO$_3$ compounds the Mn magnetic moments order in 120$^o$ arrangements. 
The magnetic frustration arises from the frustration of the first neighbor interactions between the Mn$^{3+}$ spins in the triangular lattice (see Fig. \ref{Fig1}). Yb moments of the 4b sites are antiferromagnetically coupled within a given layer. The c and c=+1/2 layers are ferromagnetically coupled.  Yb atoms of 2a sites order below 3.5 K with much smaller moments.\cite{Fabreges2008, Fiebig2003} It has been shown that the Yb(4b) moments order due to the Mn molecular field whereas the Yb(2a) moments order through Yb-Yb interactions.\cite{Fabreges2008}

The vibrational modes detected by Raman spectroscopy depend on   
the crystal symmetry which controls the matrix elements of  the Raman tensor and on  
the incident and scattered light polarizations which stress the Raman tensor. 
The irreductible representation for the normal modes in this system gives 60 phonon modes at the $\Gamma$-point: $10A_1+15A_2+10B_1+5B_2+15E_1+15E_2$ and 38 of these modes are Raman-active: $\Gamma_{Raman} = 9A_1+14E_1+15E_2$.
All measurements have been performed in backscattering configuration 
(incident wave vector anti-parallel to the scattered one). 
Pure E$_1$ and E$_2$ modes are obtained using y(zx)\={y} and z(xy)\={z} geometries, respectively.\cite{Porto1966}
The A$_1$ modes are deduced from parallel polarizations. The y(zz)\={y} and z(xx)\={z} configurations give the A$_1$(TO) + E$_2$ and A$_1$(LO) + E$_2$ modes, respectively.   
A LO-TO splitting is expected for the A$_1$ and E$_1$ modes, as these modes may induce a non-zero dipole moment both parallel and perpendicular to the phonon propagation direction. 

\begin{figure}
 \includegraphics*[width=9cm]{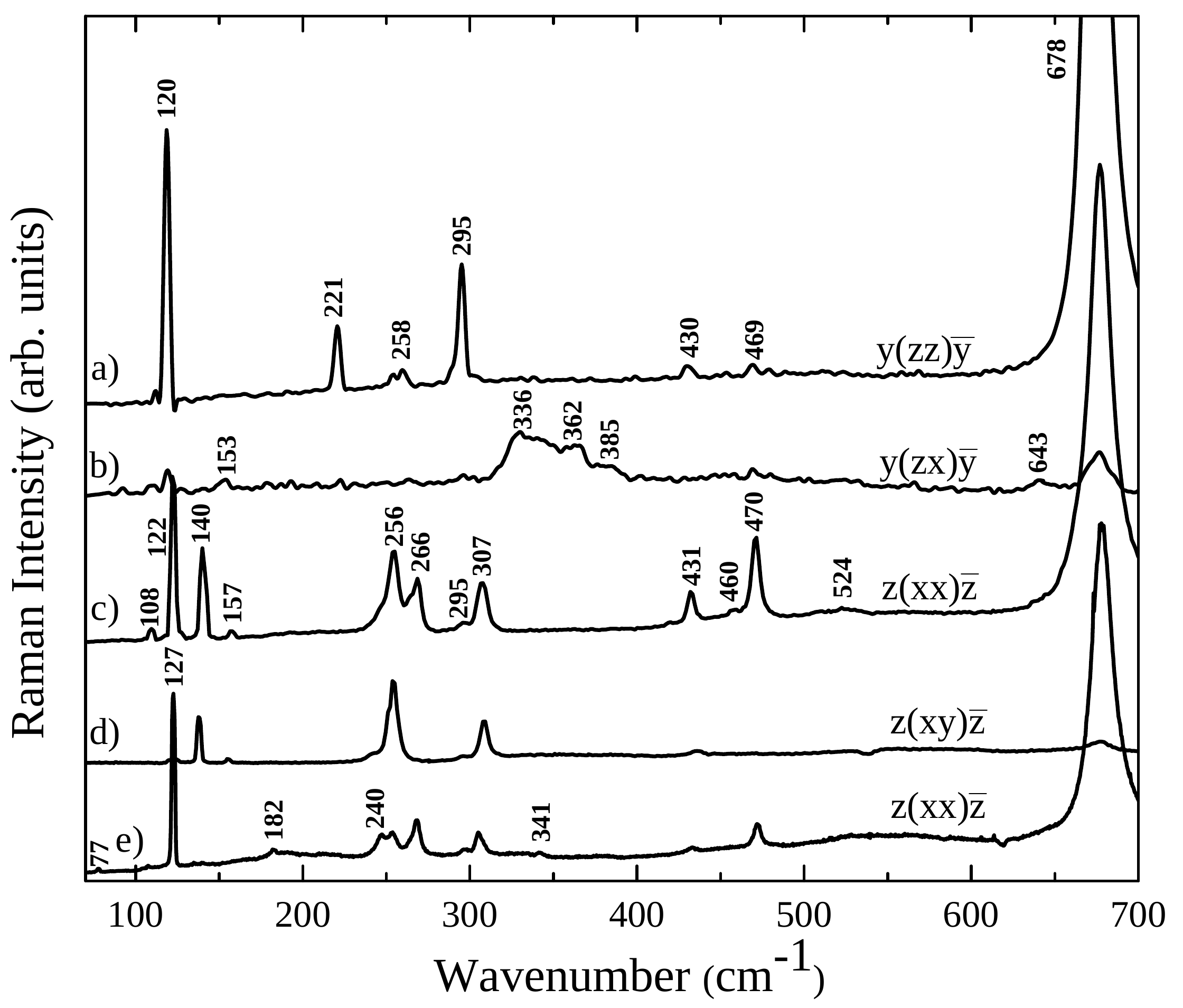} 
 \caption{\label{Fig2} 	 
Raman spectra of YbMnO$_3$ single crystal measured at 10 K using  a) y(zz)\={y} (A$_1$(TO) + E$_2$ modes),  b) y(zx)\={y} (E$_1$ modes), c) z(xx)\={z} (A$_1$(LO) + E$_2$ modes), d) z(xy)\={z} (E$_2$ modes) and e) z(xx)\={z} (A$_1$(LO) + E$_2$ modes) scattering configurations. The a, b, c and d spectra have been measured using the 514 nm laser line, the e spectrum using the 647nm laser line.}
\end{figure}

Figure \ref{Fig2} shows the Raman spectra measured on h-YbMnO$_3$ single crystals with y(zz)\={y}, y(zx)\={y}, z(xx)\={z}, and z(xy)\={z} scattering configurations. Following the Raman selection rules, we have identified 9 A$_1$ modes, 5 E$_1$ modes and 11 E$_2$ modes. 
To our knowledge the phonon modes of h-YbMnO$_3$ have been only reported in thin films and in polycrystals.\cite{Fukumura2007, Fukumura2009}  
The frequencies of the phonon modes at 10~K are reported in Table I and compared to the previous experimental results. 
The attribution of the direction and sign of atomic displacements in Table I has been based on previous works on the hexagonal RMnO$_3$ compounds.\cite {Iliev1997, Litvinchuk2004} 
We have measured a small LO-TO splitting for A$_1$ modes as expected in this manganite family.

In Fig. \ref{Fig2} we can notice that the intensity of the A$_1$ mode at 678 cm$^{-1}$ is very strong compared to the other modes. This mode is related to the motion of apical oxygen along the polarization direction (c axis) and is directly connected to the polarization of YbMnO$_3$. In order to analyze the effects of the phase transitions on the phonon modes, one can follow the temperature dependences. 

\begin{figure}
 \includegraphics*[width=9cm]{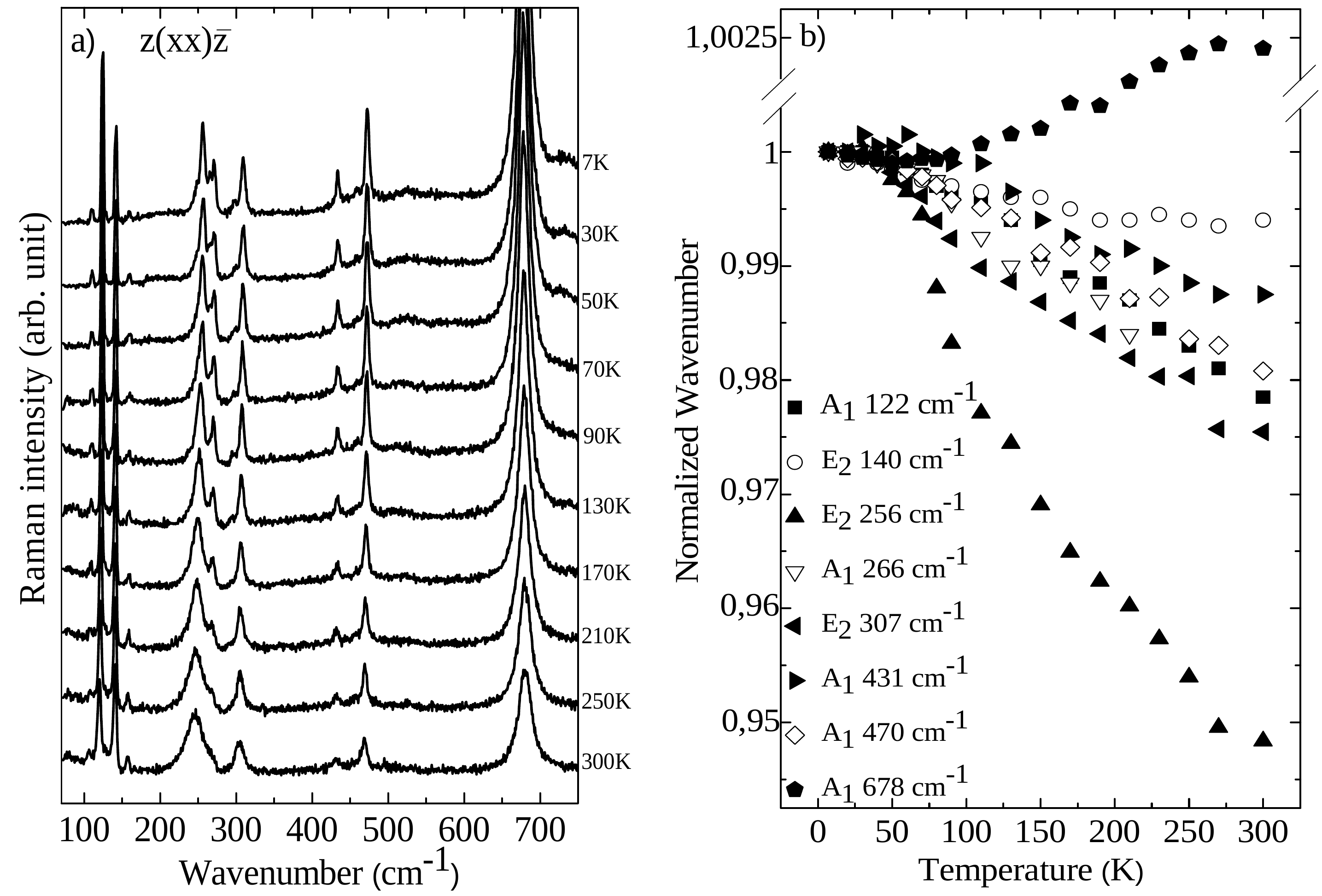} 
 \caption{\label{Fig3} 	 
a) Raman spectra of h-YbMnO$_3$ from 7 to 300 K in z(xx)\={z} configuration b) Normalized wavenumbers ($\omega$(T)/$\omega$(7K)) of 5 A$_1$ and 3 E$_2$ modes as a function of temperature.}
\end{figure}

\begin{figure}
 \includegraphics*[width=6cm]{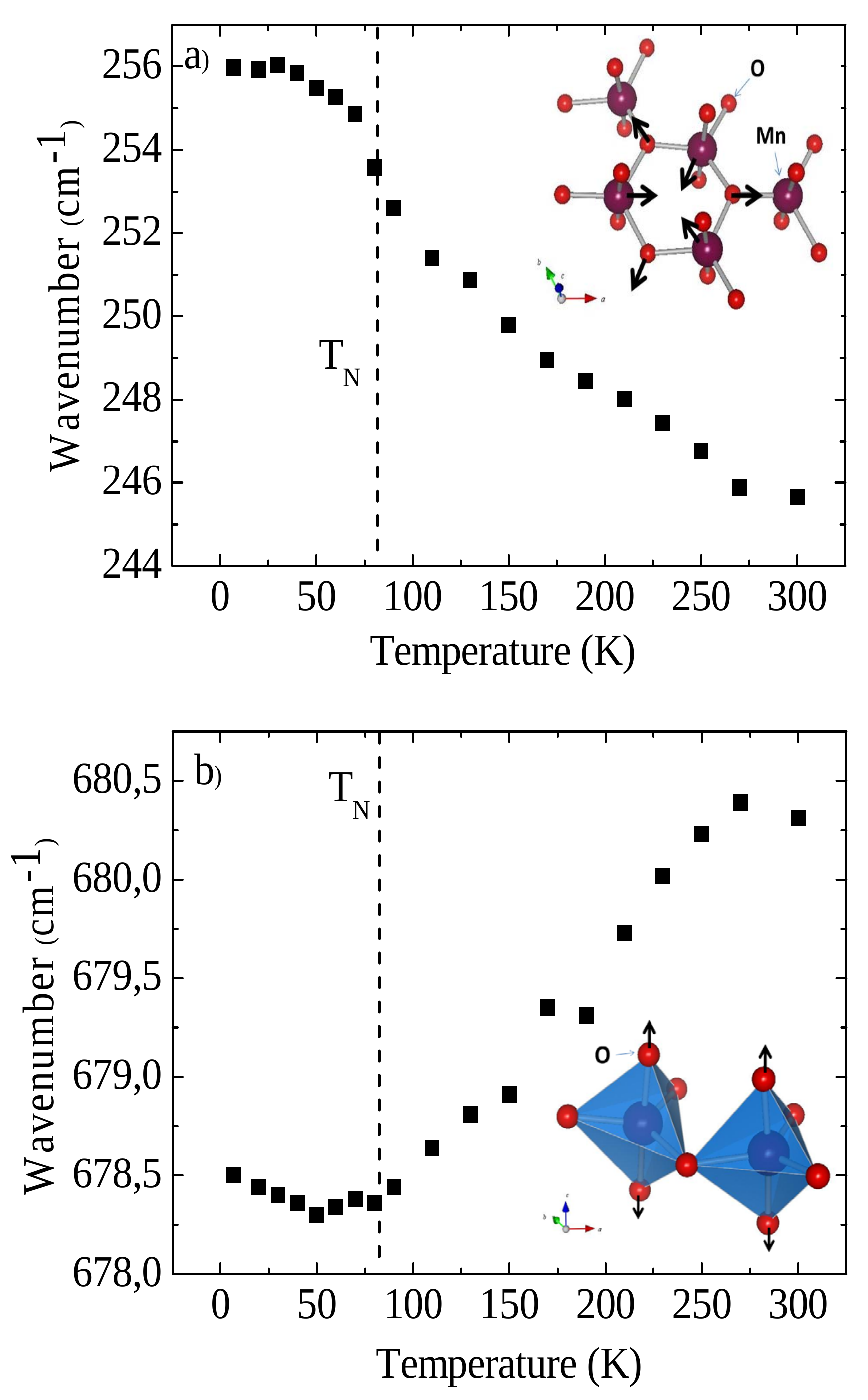} 
 \caption{\label{Fig4} 	 
Wavenumbers of the a) E$_2$ mode at 256 cm$^{-1}$  and b) A$_1$ mode at 678 cm$^{-1}$ as a function of temperature.}
\end{figure}

Figure \ref{Fig3}(a) shows the temperature dependance of Raman spectra and Fig.~\ref{Fig3}(b) represents the normalized frequencies over the frequency at 7 K of several A$_1$ and E$_2$ modes.
The phonon frequencies usually tend to soften due to the dilation of the unit cell when temperature increases. 
Except for the A$_1$ mode at 678 cm$^{-1}$, all frequencies are higher at low temperatures. In addition to this mode, the E$_2$ mode at 256 cm$^{-1}$ presents a frequency shift beyond the mean behavior of the other modes (Fig.~\ref{Fig3}(b)). The frequencies of the both modes as a function of the temperature are showed in Fig.~\ref{Fig4}. 

\begin{table}[ht]
  \begin{center}
  \parbox{8cm}
  \caption{ 
  
  A$_1$, E$_1$ and E$_2$ mode frequencies (cm$^{-1}$) measured in h-YbMnO$_3$ and description of the atomic displacements.}\\
	\vspace{3mm}
	 \begin{tabular}{cccc}
	\hline\noalign{\smallskip}
	    Mode & Fukumura & This work & Direction of the largest \\
	       & et al.\cite{Fukumura2007} & (LO/TO) & largest displacement\cite{}\\
	\hline\noalign{\smallskip}
	  A$_1$ & - & 109/-- & +Z(Yb$_1$), -Z(Yb$_2$)  \\
	        & 120 & 122/120 &  Rot x,y (MnO$_5$)    \\
	        & -  & 191/-- & +Z(Yb$_1$, Yb$_2$), -Z(Mn)  \\
	        & - & 247/221 & X(Mn), Z(O$_3$)  \\
	        & 262  & 266/258 & +Z(O$_3$),- Z(O$_4$)  \\
	        &   &  & +X,Y(O$_2$), -X,Y(O$_1$)  \\
	        & 432  & 432/430 & + Z(O$_4$,O$_3$), -Z(Mn) \\
	        & 470  & 470/469 & +X,Y(O$_1$, O$_2$), -X,Y(Mn)  \\
	        & 520  & 524/-- & + Z(O$_1$,O$_2$), -Z(Mn)  \\
	        & 677 & 678/678 & + Z(O$_1$), -Z(O$_2$)  \\
	  \hline\noalign{\smallskip}
	  E$_{1}$  & - & 152 & +X,Y(Yb$_1$), -X,Y(Yb$_2$) \\
	  				 & - & 336 & +X,Y(O$_1$, O$_2$, O$_3$) \\
	  				 &   &  &  -X,Y(O$_4$, Mn)  \\
	  				 & 360  & 362 & +X,Y(O$_1$), -X,Y(O$_2$)\\
	  				 & -  & 384 & +X,Y(O$_1$), -X,Y(O$_2$) \\
	  				 &  - & 643 & X,Y(O$_3$), -X,Y(O$_4$) \\
	  \hline\noalign{\smallskip}
	  E$_{2}$  & - & 77.5 & X,Y(Yb$_1$, Yb$_2$, Mn) \\
	           & - & 127 & +X,Y(Mn,O$_4$,O$_3$) \\
	           &   &  &  -X,Y(Yb$_1$, Yb$_2$)  \\
	           & 139 & 140 & +X,Y(Yb$_1$), -X,Y(Yb$_2$) \\
	           & 156 & 157 & +X,Y(Yb$_2$), -X,Y(Yb$_1$) \\
	           & - & 182 & -  \\
	           & 252  & 256 & +X,Y(Mn), -X,Y(O$_2$,O$_3$) \\
	           & -  & 296 & Z(Mn,O$_1$,O$_2$) \\
	           & 305 & 307 & Z(O$_1$, O$_2$), +X,Y(O$_4$) \\
	           & -  & 341 & +X,Y(O$_1$, O$_2$, O$_3$, O$_4$) \\
	           &   &  &  -X,Y(Mn)  \\
	           & -  & 419 & +X,Y(O$_1$, O$_4$), -X,Y(O$_1$, Mn) \\
	           & -  & 457.5 & +X,Y(O$_4$), -X,Y(O$_1$ , Mn) \\
	  \hline\noalign{\smallskip}
	 \end{tabular}
	\end{center}
	\label{freqphonon}
 \end{table}
 
The frequency shift of E$_2$ mode at 256 cm$^{-1}$ presents an abrupt change of slope around the N\'eel temperature in Fig.~\ref{Fig4}(a). 
Similar anomalies have been observed in the thin film and polycrystal samples of YbMnO$_3$ \cite{Fukumura2007, Fukumura2009} and in the isomorphic compounds, HoMnO$_3$ and YMnO$_3$.\cite{Vermette2008,Vermette2010}
This mode is associated to the relative displacement of Mn and O ions in the a-b plane. 
It modulates the Mn-O-Mn bond angles in a-b plane and the inplane Mn-Mn superexchange interaction. Therefore, the hardening of the E$_2$ mode frequency below 80 K characterizes the spin-phonon coupling in the magnetically-ordered phase.

The frequency of the A$_1$ mode at 678 cm$^{-1}$ in Fig~\ref{Fig4}(b) presents a softening at the magnetic transition followed by a hardening at lower temperature.
This measurement is different from the observations on YbMnO$_3$ polycrystal samples in which the frequency of this mode is constant down to 150 K and after present a softening down to 10 K without relevant changes around T$_N$.
This mode is related to the relative displacement of the apical oxygen ions along the c direction. 
It modulates the Mn-O-O-Mn bond angles, and, hence, the Mn-Mn interaction between the adjacent Mn planes. 
These interaction is due to the super-superexchange paths by the way of the apical oxygen and leads to the three dimensional magnetic ordering below T$_N$. 
The partial softening of the A$_1$ mode underlines the structural link between the ferroelectric and magnetic order in this compound. 

\subsection{Spin excitations}

\begin{figure}
 \includegraphics*[width=8cm]{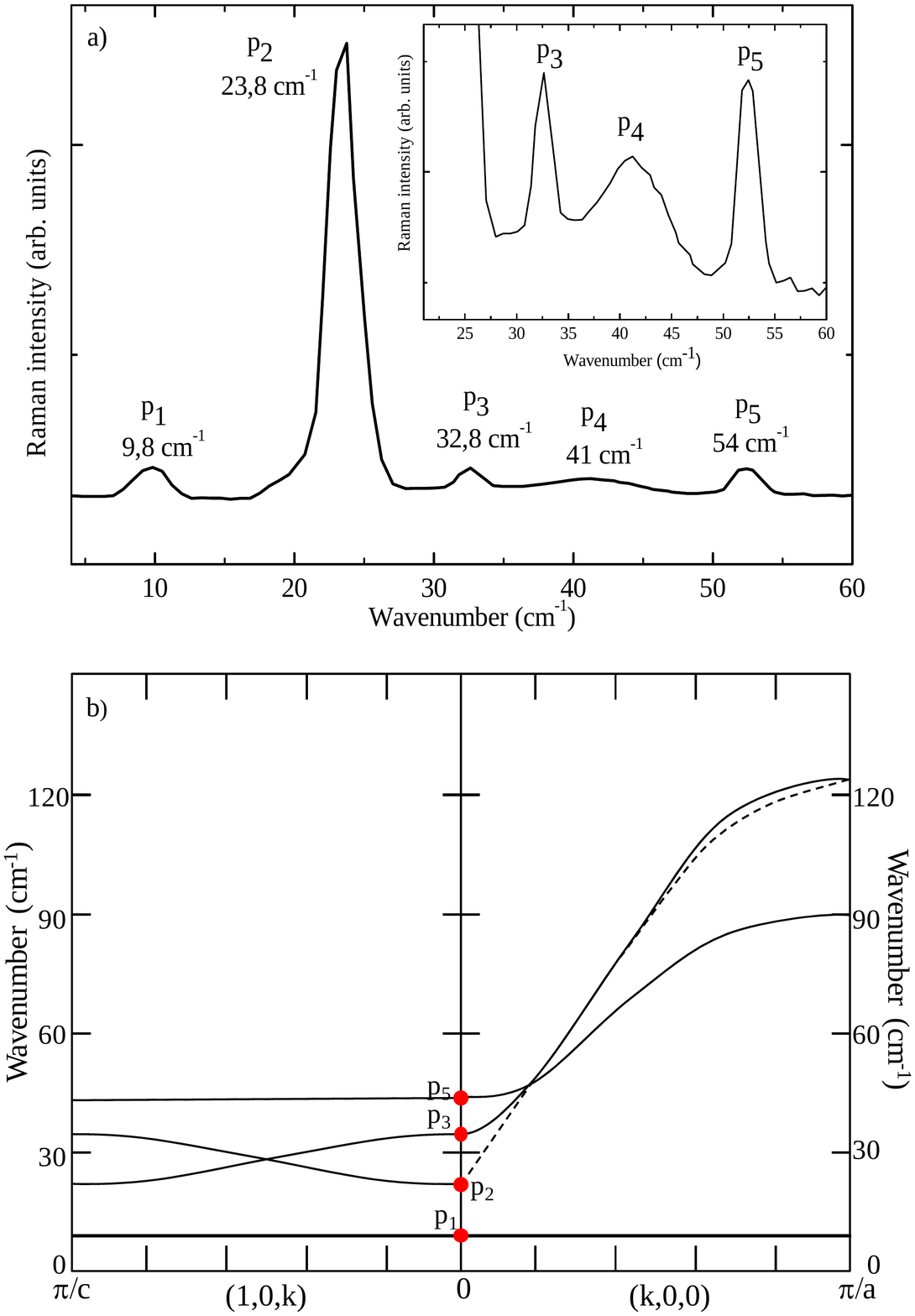} 
 \caption{\label{Fig5} 	 
a) Raman spectra of low frequency excitations measured at 7 K in z(xy)$\bar{z}$ (or y(zx)$\bar{y}$) configuration, b) crystal field and magnon dispersions along \textbf{k}= (1,0,k) and \textbf{k}= (k,0,0) extracted from Ref. \onlinecite{Fabreges2009, Fabreges2010}. The dashed lines represent the dispersion curves predicted by the theory.}
\end{figure}

Figure~\ref{Fig5}(a) shows the low frequency Raman spectra measured on h-YbMnO$_3$ single crystal at 7 K. Five peaks are detected: P$_1$ = 9.8 cm$^{-1}$, P$_2$ = 23.8 cm$^{-1}$, P$_3$ = 32.8 cm$^{-1}$, P$_4$ = 41 cm$^{-1}$ and P$_5$ = 53 cm$^{-1}$. The inset in Fig.~\ref{Fig5}(a) is a zoom in the region of the P$_3$, P$_4$ and P$_5$ peaks. 
No phonon mode is expected under 70 cm$^{-1}$. Raman scattering probes zone center (one magnons process), zone edge (two-magnon process) spin excitations and crystal field excitations. These peaks can be thus attributed to spin excitations i.e. magnon modes or to crystal field excitations. 
The origin of the excitations is discussed below based on neutron measurements. 
Figure \ref{Fig5}(b) represents the spin excitation dispersions along  the \textbf{k}= (1,0,k) and \textbf{k}= (k,0,0) axes.\cite{Fabreges2008}
The non-dispersive curve at 9.3 cm$^{-1}$ corresponds to the splitting of the fundamental doublet of the Yb(4b) ions crystal field in the molecular field of Mn ions. 
The Yb(4b) ions have a odd number of electrons on its outer layer and thus a double degeneracy of the fundamental state.\cite{Kramers1936} The molecular field of the Mn ions creates a small Zeeman splitting and the transition can be observed. 
The flat branches sitting at 24.8 and 30.6 cm$^{-1}$ are associated to the magnon modes of the Mn$^{3+}$ magnetic structure in the (a,b) plane. 
At the $\Gamma$-point, the both modes correspond to the global in phase and out of phase rotations of the 120$^o$ pattern inside the basal plane (see Fig. \ref{Fig7}(a)). 
The mode at 24.8 cm$^{-1}$ corresponds to the uniaxial anisotropy gap. The mode at 54 cm$^{-1}$ has been  associated to the Mn$^{3+}$ spins in Ref. \onlinecite{Fabreges2009} and \onlinecite{Fabreges2010}. This dispersion curve is quadruply degenerated along the c axis.
Comparing the Raman and neutron measurements of Fig.~\ref{Fig5}(a) and (b), the P$_1$ Raman peak can be clearly identified to the crystal field excitation at 9.3 cm$^{-1}$ and the P$_2$, P$_3$ and P$_5$ peaks correspond to one-magnon excitations at the center of the Brillouin zone at 24.8, 30.6 and 54 cm$^{-1}$, respectively. 

\begin{figure}
 \includegraphics*[width=8cm]{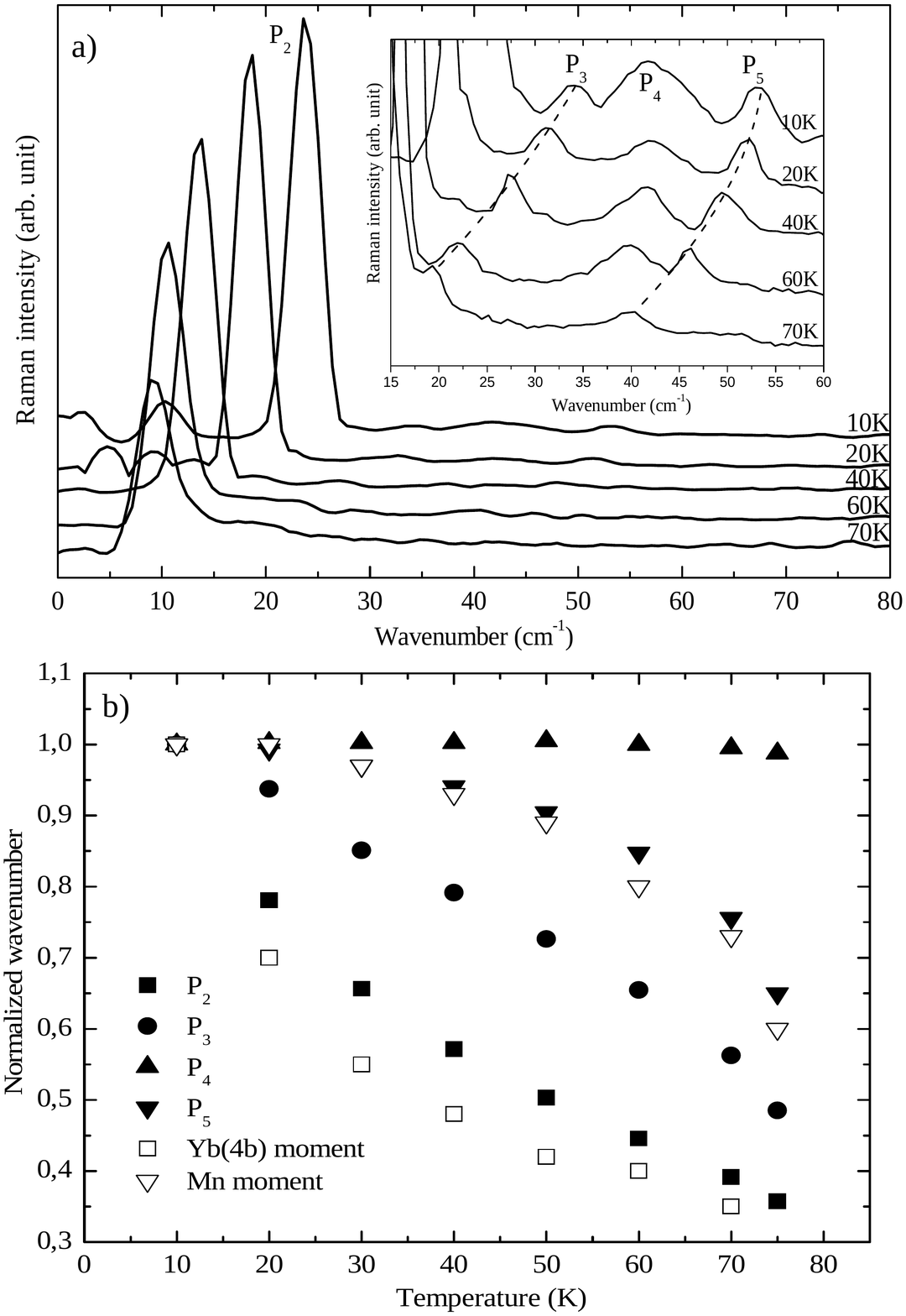} 
 \caption{\label{Fig6} 	 
a) Temperature-dependent Raman spectra in the low frequency region between 7 and 70 K b) Temperature dependence of normalized wavenumber of magnetic modes compared with the Yb and Mn moments. The Yb and Mn moment datas combine neutron powder diffraction and Mossbauer spectroscopy measurements.\cite{Fabreges2009}}
\end{figure}

The temperature dependence of the spin excitations recorded from 7 to 70 K in the z(xy)\=z configuration is shown in Fig. \ref{Fig6}(a).
The four P$_2$, P$_3$, P$_4$ and P$_5$ peaks disappear at the N\'eel temperature (80 K) and are thus related to the magnetic phase. The peaks exhibit a conventional softening as the temperature is decreased. Only the P$_4$ peak presents a striking behavior with a constant frequency with decreasing the temperature. This excitation is discussed below. Figure~\ref{Fig6}(b) shows the thermal evolution of the normalized frequency of the these peaks compared to the one of the Yb and Mn moments.\cite{Fabreges2008} The frequency shift of the P$_5$ (53 cm$^{-1}$) peak agrees well with the variation of the Mn moment. This magnon mode can be thus associated to the spin excitations of the Mn$^{3+}$ ions. The P$_2$ peak frequency shift is close to the behavior of the Yb moment. The P$_2$ peak is the signature of the uniaxial anisotropy gap in the (a,b) plane and varies with temperature similar to the Yb(4b)$^{3+}$ ion moment. This evidence points out the role of the Yb(4b)$^{3+}$ ion moment and of the Yb-Mn interaction on the P$_2$ spin wave. The spin excitation P$_3$ in Fig. \ref{Fig6}(b) presents an intermediate temperature behavior between the Yb and Mn moments that reinforce the idea of the Yb-Mn coupling. The spin excitations reflect the magnetic structure and the underlying interactions.  Those measurements show that the Mn spin orientation in the (a, b) plane is strongly influence by the Yb-Mn interaction.       
 
To shed some light on the impact of the magnetic structure on the spin excitations, we have investigated the magnetic phase diagram of h-YbMnO$_3$. 
Figure \ref{Fig7}(b) shows low frequency Raman spectra under a magnetic field $\textbf{B}$//$c$. 
The frequencies of the magnetic excitations are reported as a function of the applied magnetic field in Fig. \ref{Fig7}(c). The frequency of the P$_2$ excitation is constant, whereas the P$_3$ mode frequency decreases above 2 T and presents a crossover with the P$_2$ mode frequency around 7 T. The P$_4$ magnetic excitation disappears at 2 T. The P$_5$ peak splits at 2 T, where one component increases in frequency and the other deceases. Second harmonic generation measurements have shown a magnetic transition at 10 K for 2 T from the spin configuration B$_2$ to the configuration A$_2$ (Fig.\ref{Fig7}(a)).\cite{Fiebig2003} Notice that no hysteresis region in the phase diagram has been identified in our measurements. 
The B$_2$ and A$_2$ configurations represent a triangular antiferromagnetic ordering with antiferromagnetic and ferromagnetic coupling between adjacent planes along the c axis, respectively.  
The P$_5$ magnon mode corresponds to the Mn$^{3+}$ spin excitations that are fourly degenerated along the c axis.\cite{Fabreges2008} The Mn$^{3+}$ spin reorientation from the B$_2$ to the A$_2$ configuration partially overcomes the degeneracy of this excitation. Notice that, if the P$_5$ peak at 54 cm$^{-1}$ is not degenerate at 0 T, the maximal gap between the non degenerate curves (Raman frequency resolution) is about 0.75 cm$^{-1}$ (0.1 meV).
The second magnon mode labelled P$_3$ is a spin excitation influenced by the Yb-Mn interaction as shown by the temperature measurements in Fig. \ref{Fig6}(b).\cite{Pailhes2009} It has been shown in ErMnO$_3$ that the magnetic phase diagram is driven by the Er-Mn interaction.\cite{FiebigPRL2002} The strong impact of the magnetic field on the P$_3$ mode shows that the Mn-O-Yb superexchange is mostly involved in the B$_2$$\rightarrow$A$_2$ transition.
 The dominant magnetic interaction is the Mn-O-Mn  antiferromagnetic superexchange within the planes whereas the Mn-O-O-Mn superexchange between neighboring planes is weaker by 2 orders of magnitude. The fact that the A$_1$ phonon mode at 678 cm$^{-1}$ presents a softening due to the interplane Mn-Mn interaction and the role of the Mn-O-Yb interaction for the magnetic transition show that the weaker exchange interaction drives the magnetic transitions in h-YbMnO$_3$.

\begin{figure}
 \includegraphics*[width=9cm]{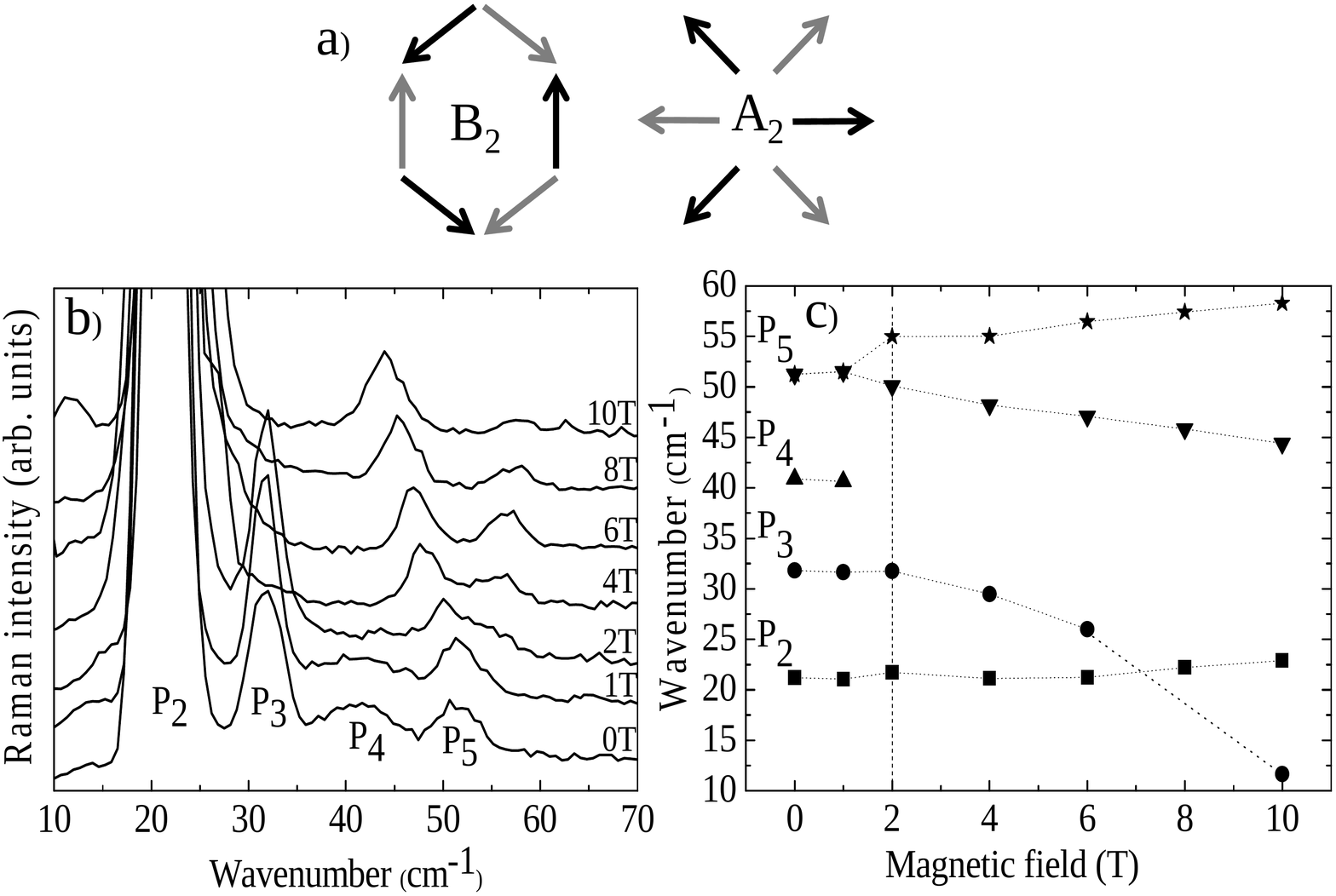} 
 \caption{\label{Fig7} 	 
a) Antiferromagnetic ordering B$_2$ and A$_2$ in h-YbMnO$_3$. Black and grey arrows indicates the Mn spin direction in the c=0 and c=1/2 planes, respectively  b) Raman spectra obtained at 10 K in z(xy)$\bar{z}$ configuration using a magnetic field along the c axis c) Wavenumber of P$_2$, P$_3$, P$_4$, P$_5$ excitations as a function of the magnetic field.}
\end{figure}

Let us focus on the origin of the P$_4$ peak at 41 cm$^{-1}$. 
The P$_4$ peak only exists below the N\'eel temperature (Fig. \ref{Fig6}(a)) and thus related to a magnetic excitation.
Based on neutron measurements, this magnetic excitation does not correspond to a one magnon mode at the center of the Brillouin zone (Fig. \ref{Fig5}(b)). However an other mode on the dispersion curves can be proposed. The P$_4$ peak can correspond to a two magnon excitation with twice the energy (around 23 cm$^{-1}$) of the zone edge along the (1,0,k) direction of the P$_3$ excitation. 
The experimental frequency of the two-magnon maximum is not exactly equal to twice the zone-edge magnon frequency, but is slightly lower in frequency
due to interaction between pairs of spin waves created close together.
The width of this peak is larger (Fig. \ref{Fig5}(a)) than the other magnetic excitations and the peaks appears only using cross polarizations as expected for a second order scattering process. The zone-edge excited magnons are short in wavelength
and are thus sensitive to short-range magnetic order. It explains the reduced sensitivity of its frequency to the temperature (Fig. \ref{Fig6}(b)). Moreover, this peak disappears in the A$_2$ magnetic phase as it shown in Fig. \ref{Fig7}(a).  Knowing that the intensity of the two-magnon excitations is  essentially proportional to the two-magnon density of states, the transition from the B$_2$ to the A$_2$ magnetic phase renormalizes strongly the zone edge density of states of the associated magnon mode. 

\section{conclusion}
In summary, our measurements show that the A$_1$ phonon mode related to the displacement of the apical oxygen ions and connected to the macroscopic polarization is in strong interaction with the magnetic sublattice by the way of the Mn-O-O-Mn interaction. We also reveal the spin excitations and the role played by the interplane Mn-O-Yb superexchange interaction that drives the magnetic transitions. 

\section*{Acknowledgments}
This work was supported in part by the Australian Research council (grant DP-110105346).
The work at Rutgers University was supported by NSF-DMR-1104484.


\begin{thebibliography}{00}

\bibitem{Yang2007} F. Yang, M. H. Tang, Z. Ye, Y. C. Zhou, X. J. Zheng, J. X. Tang, J. J. Zhang, and J. He, J. Appl. Phys. {\bf 102}, 044504 (2007).

\bibitem{Fiebig2002} M. Fiebig, T. Lottermoser, D. Fr$\ddot{ö}$hlich, A. V. Goltsev, and R. V. Pisarev, Nature London {\bf 419}, 818 (2002).

\bibitem{Moussa1996} F. Moussa, M. Hennion, J. Rodriguez-Carvajal, H. Moudden, L. Pinsard, and A. Revcolevschi, Phys. Rev. B {\bf 54}, 15149 (1996).

\bibitem{Tokura2006} Y. Tokura, Science {\bf 312}, 1481 (2006).

\bibitem{Eerenstein2006} W. Eerenstein, N. D. Mathur, and J. F. Scott, Nature (London) {\bf 442}, 759 (2006).

\bibitem{Chu2008} Y. H. Chu, L. W. Martin, M. B. Holcomb, M. Gajek, S-J. Han, Q. He, N. Balke, C-H. Yang, D. Lee, W. Hu, Q. Zhan, P-L. Yang, A. Fraile-Rodríguez, A. Scholl, S. X. Wang and R. Ramesh, Nature Mater. {\bf 7}, 478 (2008).

\bibitem{Baek2010} S. H. Baek, H. W. Jang, C. M. Folkman, Y. L. Li, B. Winchester, J. X. Zhang, Q. He, Y. H. Chu, C. T. Nelson, M. S. Rzchowski, X. Q. Pan, R. Ramesh, L. Q. Chen, and C. B. Eom, Nature Mater. {\bf 9}, 309 (2010).

\bibitem{Fiebig2005} M. Fiebig, J. Phys. D {\bf 38}, R123-R150 (2005).

\bibitem{Cheong2007} S. W. Cheong and M. Mostovoy, Nature Mater. {\bf 6}, 13 (2007).

\bibitem{Park2003} J. Park, J. G. Park, G. S. Jeon, H. Y. Choi, C. H. Lee, W. Jo, R. Bewley, K. A. McEwen, and T. G. Perring, Phys. Rev. B {\bf 68}, 104426 (2003).

\bibitem{Lee2008} S. Lee, A. Pirogov, M. Kang, K. H. Jang, M. Yonemura, T. Kamiyama, S.-W. Cheong, F. Gozzo, N. Shin, H. Kimura, Y. Noda, and J. G. Park, Nature (London) {\bf 451}, 805 (2008).

\bibitem{Pimenov2006} A. Pimenov, T. Rudolf, F. Mayr, A. Loidl, A. A. Mukhin, and A. M. Balbashov, Phys. Rev. B {\bf 74}, 100403 (R) (2006).

\bibitem{Aken2004} B. B. Van Aken, T. T. M. Palstra, A. Filippetti, and N. A. Spaldin, Nature Mater. {\bf 3}, 164 (2004).

\bibitem{Divis2008} M. Divis, J. Holsa, M. Lastusaari, A. P. Litvinchuk and V. Nekvasil J. Alloys Comp. {\bf 451}, 662 (2008).

\bibitem{Fabreges2008} X. Fabr\'eges, I. Mirebeau, P. Bonville, S. Petit, G. Lebras-Jasmin, A. Forget, G. Andre, and S. Pailhes,  Phys. Rev. B {\bf 78}, 214422 (2008).

\bibitem{Fabreges2009} X. Fabr\'eges, S. Petit, I. Mirebeau, S. Pailhes, L. Pinsard, A. Forget, M. T. Fernandez-Diaz, and F. Porcher,  Phys. Rev. Lett. {\bf 103}, 067204 (2009).

\bibitem{Fiebig2003} M. Fiebig, Th. Lottermoser, and R. V. Pisarev, J. Appl. Phys.{\bf 93}, 8194 (2003).

\bibitem{Fukumura2007} H. Fukumura, N. Hasuike, H. Harima, K. Kisoda, K. Fukae, T. Takahashi, T. Yoshimura and N. Fujimura,  J. Phys. : Conference Series {\bf 92},  012126 (2007).

\bibitem{Fukumura2009} H. Fukumura, N. Hasuike, H. Harima, K. Kisoda, K. Fukae, T. Takahashi, T. Yoshimura and N. Fujimura,  J. Phys. : Condens. Matter {\bf 21}, 064218 (2009).

\bibitem{Kim2000} T. Choi, Y. Horibe,	H. T. Yi,	Y. J. Choi,	Weida Wu and S.-W. Cheong, Nature Mater. {\bf 9}, 253 (2010).

\bibitem{Munoz2000} A. Munoz, J. A. Alonso, M. J. Martinez-Lope, M. T. Casais, J. L. Martinez, and M. T. Fernandez-Diaz, Phys. Rev. B {\bf 62}, 9498 (2000).

\bibitem{Porto1966} S. P. S. Porto, J. A. Giordmaine, and T. C. Damen, Phys. Rev. {\bf 147}, 608 (1966).

\bibitem{Iliev1997} M. N. Iliev, H. G. Lee, V. N. Popov, M. V. Abrashev, A. Hamed, R. L. Meng and C. W. Chu, Phys. Rev. B {\bf 56}, 2488 (1997). 

\bibitem{Litvinchuk2004} A. P. Litvinchuk, M. N. Iliev, V. N. Popov and M. M. Gospodinov,  J. Phys. : Condens. Matter {\bf 16}, 809 (2004).

\bibitem{Vermette2008} J. Vermette, S. Jandl, and M. M. Gospodinov, J. Phys.: Condens. Matter {\bf 20}, 425219 (2008).

\bibitem{Vermette2010} J. Vermette, S. Jandl, A. A. Mukhin, V. Yu. Ivanov, A. Balbashov, M. M. Gospodinov, and L. Pinsard-Gaudart,  J. Phys.: Condens. Matter {\bf 22}, 356002 (2010).

\bibitem{Fabreges2010} X. Fabr\'eges, PhD Thesis, Etude des propriétés magnétiques et du couplage spin/réseau dans les composés multiferroïques RMnO$_3$ hexagonaux par diffusion de neutrons, University Paris-Sud 11, France (2010).

\bibitem{Kramers1936} H. Kramers, Proc. Acad. Sci. Amsterdam {\bf 33}, 959 (1936).

\bibitem{Pailhes2009} S. Pailh\'es,  X. Fabr\'eges,  L. P. R\`egnault, L. Pinsard-Godart, I. Mirebeau, F. Moussa, M. Hennion,  and S. Petit, Phys. Rev. B {\bf 79}, 134409 (2009). 

\bibitem{FiebigPRL2002} M. Fiebig, C. Degenhardt, and R. V. Pisarev, Phys. Rev. Lett. {\bf 88}, 027203 (2001).

\end{thebibliography}
\end{document}